# The gravitational field energy density for symmetrical and asymmetrical systems

**Roald Sosnovskiy**
Technical University, 194021, St. Petersburg, Russia
E-mail:rosov2@yandex

**Abstract.** The relativistic theory of gravitation has the considerable difficulties by description of the gravitational field energy. Pseudotensor $t_0^0$ in the some cases cannot be interpreted as energy density of the gravitational field. In [1] the approach was proposed, which allow to express the energy density of such a field through the components of a metric tensor. This approach based on the consideration of the isothermal compression of the layer consisted of the incoherent matter. It was employ to the cylindrically and spherically symmetrical static gravitational field. In presented paper the approach is developed.

**1. Introduction.** The problem of the gravitational field energy discussed a long time [2], [3]. However, pseudotensor $t_\nu^\mu$ differs from author to author reflecting the ambiguity in defining gravitational field energy density [3]. In [1] the approach has proposed allows one to express the energy density of such a field through the components of a metric tensor. This approach based on the consideration of the isothermal compression of a layer consisted of the incoherent matter in the field of the infinitesimal thin material shell by fulfillment of the requirements [4]: (a) the local energy conservation law should be fulfilled and (b) the correspondence principle should be fulfilled including the energy part.

In the presented paper proved, that this approach can be used for asymmetrical systems. Here proved, that the requirement of the invariance of the gravitational field energy density [4] fulfilled. For the cylindrically and spherically symmetrical systems is obtained field energy density formulas, contained only the metric tensor component and his derivative.

**2. The differential of the gravitational field energy**

In [1] has obtained the formulas of the gravitational field energy for the special coordinates, connected with type of symmetry. Here it considered the formula of the field energy for the arbitrary static coordinates systems. The solution is analogous to one in [1].

**2.1. The isothermal compression.** Here it considered the movement of the particles layers when acquired energy of particles has eradiated or dissipated. The movement considered as consisted of discrete infinitesimal steps, when the particles fall free, and in end of step energy of particles has dissipated. Concrete ways of dissipation no discussed. Sufficiently to suppose that such way can be on principle approximately realized. For example, free fall of particles in thin lay on the solid surface with following cooling of the solid.

The particles considered as test-particles. However, the change of field, caused with accumulation of the matter on solid surface, calculated after every step.

Assume $\bar{x}^\alpha$ is the initial coordinate system, admissible for the system configuration, with metric $\bar{g}_{\mu\nu}, \bar{g}_{\mu 0} = 0$. Let us consider the displacement of the particles layer from position $\bar{x}^1 = \bar{x}_1^1$ to position $\bar{x}^1 = \bar{x}_1^1 + d\bar{x}^1$. The free particles fall equations are [5]

$$\frac{d}{d\tau}\left(\frac{\partial L}{\partial \dot{\bar{x}}^\mu}\right) - \frac{\partial L}{\partial \bar{x}^\mu} = 0 \qquad (1)$$

where $\tau$ is the intrinsic time, $\dot{\bar{x}}^\mu = \frac{d\bar{x}^\mu}{d\tau}$ and

$$L(\bar{\dot{x}}^\sigma, \bar{x}^\sigma) = \frac{1}{2}\bar{g}_{\mu\nu}\bar{\dot{x}}^\mu\bar{\dot{x}}^\nu, \qquad \bar{g}_{\mu\nu} = \bar{g}_{\mu\nu}(\bar{x}^1, \bar{x}^2, \bar{x}^3) \qquad (2)$$

For static system (1),(2) lead to

$$\bar{\dot{x}}^0 = \frac{C_0}{\bar{g}_{00}} \qquad (3)$$

where $C_0$ is constant on all step. From (3) and from equation

$$c^2 = \bar{g}_{00}\bar{\dot{x}}^{02} + \bar{g}_{ik}\bar{\dot{x}}^i\bar{\dot{x}}^k \qquad (4)$$

we get, so far as $\bar{\dot{x}}^i$ is small,

$$C_0 = c\sqrt{\bar{g}_{00}} \qquad (5)$$

From formulas (1), (2) for i, k =1,2,3 result

$$\bar{g}_{ik}\bar{\ddot{x}}^k = \frac{1}{2}\bar{g}_{00,i}\bar{\dot{x}}^{02} \qquad (6)$$

and from (5),(6)

$$\bar{\ddot{x}}^k = \frac{1}{2}\frac{\bar{g}^{ik}\bar{g}_{00,i}c^2\delta\tau}{\bar{g}_{00}} \qquad (7)$$

where $\tau$ is the intrinsic time of particles movement. Components $\bar{u}^k$ of the maximum velocity of free particles fall near by point $\vec{\bar{x}}(\bar{x}^1, \bar{x}^2, \bar{x}^3)$ may written

$$\bar{u}^k = \beta\frac{\bar{g}^{ik}\bar{g}_{00,i}}{\bar{g}_{00}} \qquad (8)$$

where ß is infinitesimal coefficient.

**2.2. The static gravitational field energy. General formulas.** The energy of the particles by free fall can obtained from the relation [6]

$$E = \delta m c^2 \bar{u}^\nu \frac{\bar{g}_{0\nu}}{\sqrt{\bar{g}_{00}}}, \qquad \bar{u}^\nu = \frac{d\bar{x}^\nu}{cd\tau} \qquad (9)$$

where $\delta m$ is rest mass of the particles group. From (3) and so far as $\bar{g}_{0\nu} = \delta_\nu^0 \bar{g}_{00}$ the change of particles energy on way $d\bar{x}^i$ is

$$dE = -\frac{\delta m c^2}{2\bar{g}_{00}}\bar{g}_{00,i}d\bar{x}^i \qquad (10)$$

This energy has dissipated on way $d\bar{x}^i$. If the local energy conservation law fulfilled then the energy change of particles must result from change of field energy $dE_f$ on way $d\bar{x}^i$. Therefore

$$dE_f = \frac{\delta m c^2}{2\bar{g}_{00}}\bar{g}_{00,i}d\bar{x}^i \qquad (11)$$

From (8) follow, that the components of the of the particle coordinates mean change is

$$d\bar{x}^i = \lambda\frac{\bar{g}^{ik}\bar{g}_{00,k}}{\bar{g}_{00}} \qquad (12)$$

and scalar displacement is equal

$$dl = \frac{|\lambda|}{\bar{g}_{00}} \sqrt{\bar{g}^{ik} \bar{g}_{00,i} \bar{g}_{00,k}} \tag{13}$$

where

$$|\lambda| = \frac{\bar{g}_{00} dl}{\sqrt{\bar{g}^{ik} \bar{g}_{00,i} \bar{g}_{00,k}}} \tag{14}$$

If substitute $\lambda$ in (12) and then $d\bar{x}^i$ in (11) then we get

$$d\delta E_f = -\frac{\delta m c^2}{2\bar{g}_{00}} \sqrt{\bar{g}^{ik} \bar{g}_{00,i} \bar{g}_{00,k}} \, dl \tag{15}$$

Here δm and δ$l$ are scalars, $\bar{g}_{00}$ by $\bar{g}_{0i} = 0$ no depend on the space coordinates transformation. The quantity $\bar{g}^{ik} \bar{g}_{00,i} \bar{g}_{00,k}$ is invariant by the space coordinates transformation

$$\bar{x}^i = A_k^i x^k \tag{16}$$

Therefore, $d\delta E_f$ is also invariant.

### 3. Energy of the asymmetrical gravitational field.

**3.1. The object.** Considered the static field of the asymmetric convex smooth infinitesimally thin material shell with surface mass density $\sigma_c$. The quantity $\sigma_c$ is a single-valued function of the coordinates of shell point $\bar{x}_c^1 = \bar{x}_c^1(\bar{x}^2, \bar{x}^3)$, $\sigma_c = \sigma_c(\bar{x}^2, \bar{x}^3)$. It considered the space between a shell and some convex smooth external surface $\bar{x}_e^1 = \bar{x}_e^1(\bar{x}^2, \bar{x}^3)$ with surface mass density $\sigma_e = \sigma_e(\bar{x}^2, \bar{x}^3)$. Assumed, that it known how the metric $\bar{g}_{ik}$ of the space region between $\bar{x}_c$ and $\bar{x}_e$ to find. This is possible at least by miens of the computer methods [7].

**3.2. The calculations order.** Is considered the motion of $N_j$ discrete test particles layers from the external surface to the shell. The motion is discrete; the number of steps is $N_q$. For every layer position, the calculations made for $N_k \cdot N_n$ points. The every point position determined in coordinate system $\bar{x}^i$.

For every point P(k,n,q) the volume element is built at the vectors $d\bar{z}_i = d\bar{z}_i(d\bar{x}^1, d\bar{x}^2, d\bar{x}^3)$, i =1,2,3. Let this vectors create the coordinate system

$$dz^i = B_k^i(k,n,q,j) d\bar{x}^k, \quad g_{00} = \bar{g}_{00} \tag{17}$$

One side $(dz^2, dz^3)$ of this element is disposed at the layer position q and the opposite side at the layer position q+1. The vector $d\vec{l}$ describe the fall of particle from point P(k,n,q) up to point F(k',n',q+1) at the layer position q+1. For every point P(k,n,q) and every layer j are calculated from (15) the field energy differential $d\delta E_f$ and masse density σ(k',n',q+1,j) for point P(k',n',q+1). Afterwards the layer j arrive the position q = $N_q$ metric components $g_{ik}$ calculated for all points P(k,n,q). The method of such calculation no considered because that does not matter for the purpose of this paper.

**3.3. The gravitational field integral energy and energy density invariance.** Let the volume element is built at the vectors $dz^i$.

The mass of the particles group, passed through this element, is equal

$$\delta m(k,n,q,j) = \delta\sigma(k,n,q,j) \cdot dS(k,n,q,j) \tag{18}$$

where δσ(k,n,q,j) is the matter density in the particles layer j and $dS(k,n,q,j)$ – the area of the element $(dz^2, dz^3)$. The mass δm of this element, which is considered as in the one point concentrated, fall from point P(k,n,q) in point F(k,n,q+1) with coordinates $z^i + dl^i$. Components $dl^i$ can be calculated in coordinates $dz^i$ from (12), (17). Component $dl^1 = dz^1$ and

$$dl^k = dl^1 \frac{g^{kp} g_{00,p}}{g^{1q} g_{00,q}}, k=2,3 \tag{19}$$

By means interpolation can be calculate the mass δm(n,q+1,j) and mass density δσ(n,q+1,j) for points P(k,n,q+1). Consider the successive pass of the layers through the element of area $(dz^2, dz^3)$ with point P(k,n,q). From (15),(17) after step $j = N_j$ the field energy change in volume element is equal

$$dE_f(k,n,q) = \frac{c^2}{2} \sum_j^{N_j} \left[ \frac{dS \cdot \delta\sigma \cdot dl}{2g_{00}} \sqrt{g^{is} g_{00,i} g_{00,s}} \right]_{k,n,q,j} \tag{20}$$

where [] depend on (k,n,q,j). The quantities under the symbol Σ are the invariants, therefore $dE_f(k,n,q)$ is invariant. The sum of energy in all points of the field is also invariant.

The energy density in point P(n,q) is given by

$$w(k,n,q) = \frac{dE_f(n,q)}{dV(n,q)}, \quad dV = \sqrt{|g_z|} dz^1 dz^2 dz^3 \tag{21}$$

where dV(n,q) is the volume of the volume element, built at the vectors $(d\vec{z}_1, d\vec{z}_2, d\vec{z}_3)$ for step j=N_j; $|g_z|$ is determinant of the metric components. The quantity dV is scalar, therefore w(n,q) is invariant.

Thus, the approach based on the consideration of the isothermal compression of the layer consisted of the incoherent matter, can be used for asymmetrical systems.

**4. The transformation of the formulas for field energy density of the symmetrical systems.**

The formulas for these quantities in the paper [1] maintain, besides the metric tensor components, the field source mass M and the distance to symmetry centre R. As the metric tensor components are the functions of M and R, it is possible to except M and R from these formulas.

**4.1. The cylindrical symmetry.** In [1] there are the formulas

$$g_{00} = \left(\frac{R}{R_0}\right)^{a_0}, \quad a_0 = \frac{4GM_z}{c^2} \tag{22}$$

where R – radius, $R_0$ – radius of the field source, $M_z$ – the linear mass density. From (23) follow

$$\frac{g_{00,1}}{g_{00}} = \frac{4GM_z}{c^2 R} \tag{23}$$

and energy density

$$w = -\frac{GM_z^2}{2\pi R^2} \sqrt{g_{00}} = -\frac{c^4}{32\pi G} \left(\frac{g_{00,1}}{g_{00}}\right)^2 \sqrt{g_{00}} \tag{24}$$

**4.2. The spherical symmetry.** From [1] in this case

$$g_{00} = 1 - \frac{2GM}{c^2 R}; \quad dx^1 = dR; \, dx^2 = Rd\theta; \, dx^3 = R\sin\theta d\varphi \tag{25}$$

and energy density

$$w = \frac{c^2 \sqrt{g_{00}}}{16\pi G R^2} \left[ \ln\left(1 - \frac{2GM}{c^2 R}\right) + \frac{2GM}{c^2 R} \right] \tag{26}$$

or from (25)

$$w = \frac{c^4 g_{00,1}^2 \sqrt{g_{00}}}{16\pi G R^2 (1-g_{00})^2} [\ln g_{00} + 1 - g_{00}] \cong -\frac{c^4 g_{00,1}^2}{32\pi G} \qquad (27)$$